\documentclass[11pt,prd,onecolumn,showpacs,amsmath,amssymb,aps,floats,floatfix,nofootinbib]{revtex4-1}
\usepackage[colorlinks=true,urlcolor=rossos,anchorcolor=black,citecolor=mygreen,filecolor=black,linkcolor=black,menucolor=blue,pagecolor=blue,linktocpage=true]{hyperref} 

\usepackage[multidot]{grffile}  
\usepackage{dcolumn}
\usepackage{amsmath}
\usepackage{amsfonts}
\usepackage{amssymb}
\usepackage{color}
\usepackage{footmisc}
\usepackage{latexsym}
\usepackage{slashed} 
\usepackage{pstricks}
\usepackage{indentfirst}
\usepackage{mathrsfs}
\usepackage{multirow}
\usepackage{epsfig,psfrag}
\usepackage{subfigure}
\usepackage{mathtools}
\usepackage{setspace} 
\usepackage[utf8]{inputenc} 
\usepackage[scientific-notation=true]{siunitx} 
\graphicspath{{figs/}}

\def\bac{\begin{array} {c}}

\definecolor{rossos}{cmyk}{0,1,1,0.55}
\definecolor{mygreen}{rgb}{0.27, 0.64, 0.48}

\newcommand{\SU}{\text{SU}}

  \makeatother

  \allowdisplaybreaks 

\begin{document}
 	
 \title{\Large \bf \color{rossos} Exploring Fermionic Multiplet Dark Matter through Precision Measurements at the CEPC}
 	\author{Lin-Qing Gao$^{1,2}$}
 	\email{gaolq@ihep.ac.cn}
 	\author{Xiao-Jun Bi$^{1,2}$}
 	\email{bixj@ihep.ac.cn}
 	\author{Jin-Wei Wang$^{3,4,5}$}
 	\email{jinwei.wang@sissa.it}
 	\author{Qian-Fei Xiang$^6$}
 	\email{xiangqf@pku.edu.cn}
 	\author{Peng-Fei Yin$^1$}
 	\email{yinpf@ihep.ac.cn}
 	\vspace{2mm}
 	\affiliation{	\vspace{2mm}$^1$Key Laboratory of Particle Astrophysics,
 		Institute of High Energy Physics, Chinese Academy of Sciences,
 		Beijing 100049, China}
 	\affiliation{$^2$School of Physical Sciences,
 		University of Chinese Academy of Sciences,
 		Beijing 100049, China}
 	\affiliation{$^3$Scuola Internazionale Superiore di Studi Avanzati (SISSA), via Bonomea 265, 34136 Trieste, Italy}
 	\affiliation{$^4$INFN, Sezione di Trieste, via Valerio 2, 34127 Trieste, Italy}
 	\affiliation{$^5$
 		Institute for Fundamental Physics of the Universe (IFPU), via Beirut 2, 34151 Trieste, Italy}
 	\affiliation{$^6$Center for High Energy Physics, Peking University, Beijing 100871, China}

 	\begin{abstract}
 		\vspace{0.4cm}
 		\large
 		
New physics could be explored through loop effects by the precision measurements at the Circular Electron Positron Collider due to its clean collision environment and high luminosity. In this work, we focus on two dark matter models that involve additional electroweak fermionic multiplets. We calculate their one-loop corrections to five processes, i.e. $e^+e^- \to \mu^+\mu^-, ~Zh, ~ZZ, ~W^+W^-$, and $Z\gamma$, and investigate the corresponding signatures at CEPC with the projected sensitivity.
We find that the detectable parameter regions of these processes are  complementary. 
The combined analysis shows that the mass of dark matter $m_{\chi^0_1}$ in these two models can be probed up to $ \sim 150~\mathrm{GeV}$ and $\sim 450$ GeV at a 95\% confidence level, respectively.
 		
 	\end{abstract}
 	
 	\maketitle
 	\tableofcontents
 	
 	\section{Introduction}
 	\label{sec:intro}
 The discovery of Higgs boson at the Large Hadron Collider (LHC) marks a huge success for the standard model (SM) of particle physics \cite{Aad:2012tfa,Chatrchyan:2012xdj}.  
 However, the existence of dark matter (DM), which is revealed by many astrophysical and cosmological observations, clearly shows that the SM is not a complete theory and the new physics (NP) beyond SM must exist \cite{Feng:2010gw,Bertone:2004pz}. 
 Under the assumption that DM is an unknown particle, its properties could be effectively detected at colliders, especially when its mass is smaller than the center of mass energy of the collider. 
 If this is not the case, 
the properties of DM might also be investigated through the loop effects at the proposed electron-positron colliders due to their high precision, such as the Circular Electron Positron Collider (CEPC) \cite{CEPC:CDR} 
the Future Circular Collider \cite{FCC-ee:CDR}, and the International Linear Collider \cite{ILC:CDR}. 
 
 Among various DM candidates proposed in the literature, weakly interacting
 massive particles (WIMPs) are very compelling, because they can naturally explain the DM relic
 density (dubbed WIMP miracle). 
 It is natural to construct a WIMP model by introducing a dark sector contains electroweak (EW)
 $\SU(2)_L$ multiplets, such as the minimal dark matter model \cite{Cirelli:2005uq}, which involves one nontrivial $\SU(2)_L$ multiplet and is regarded as the minimal extension.
 In this work, we focus on a type of DM model that contains more than one EW multiplet:
 \begin{itemize}
 	\item Singlet-doublet fermionic dark matter (SDFDM) model: the dark sector involves one singlet Weyl spinor and two doublet Weyl spinors  \cite{Yaguna:2015mva,Calibbi:2015nha,Cai:2016sjz,Xiang:2017yfs,Wang:2018};
 	\item Doublet-triplet fermionic dark matter (DTFDM) model: the dark sector involves two doublet Weyl spinors and one triplet Weyl spinor \cite{Dedes:2014hga,Cai:2016sjz,Xiang:2017yfs,Wang:2018}.
 \end{itemize}
 Given that the even dimensional representation of the $SU(2)$ group is pseudoreal, so two doublet Weyl spinors are introduced to generate the corresponding mass terms.
 After electroweak symmetry breaking (EWSB), the Yukawa terms lead to mixing between these multiplets. Specifically, there are three neutral Majorana fermions and one (two) charged fermions in the SDFDM (DTFDM) model. With a discrete $Z_2$ symmetry, the lightest Majorana fermion is stable and can be the DM candidate.
 Considering the possible couplings between dark sector particles and the gauge bosons (e.g. $W$, $Z$, and $\gamma$) or Higgs boson, CEPC is an ideal tool to explore these models because of its accurate measurement and large luminosity.
 
 According to the CEPC operation plan \cite{CEPC:CDR}, it will work in three possible modes, including the Higgs factory, $Z$ factory, and $WW$ threshold scan. For the Higgs factory period, the CEPC will run at $\sqrt{s}=240$ GeV for  $e^+ e^- \to Z h$ production, and the total integrated luminosity can reach $\sim 5.6~\text{ab}^{-1}$ \cite{CEPC:CDR}. 
 As a result, the CEPC can collect $\sim 10^6$ Higgs bosons, $\sim 10^7$ $\mu^+\mu^-$ events, $\sim 3 \times 10^6$ $ZZ$ events, $\sim 5 \times 10^7$ $W^+ W^-$ events, and $\sim5\times 10^7$ $Z\gamma$ events. 
 With so many events, the sensitivities of these channels at CEPC could reach a sub-percentage level. Therefore, the hint of NP could be detected through the loop order processes \cite{McCullough:2013rea,Cao:2014ita,Beneke:2014sba,Shen:2015pha,Huang:2015izx,Kobakhidze:2016mfx,Xiang:2017yfs,Wang:2017sxx}. 
 Some recent works have investigated such effects in the Higgs decay \cite{Chen:2019pkq,Susufang:2018shg}, EW oblique parameters \cite{Cai:2016sjz,Cai:2017wdu}, $e^+ e^- \to W^+ W^-$ \cite{Wulei:2017kgr,Andreev:2012cj}, $e^+ e^- \to \mu^+ \mu^-$ \cite{Harigaya_2015,Cao_2016,Gulov:2013kpa}, and so on.

 In our work, we investigate the loop effects of the SDFDM and DTFDM models at the CEPC with $\sqrt{s}=240$ GeV through five processes\footnote{Note that the signature of $e^+e^- \to Zh$ has been studied in Ref. \cite{Xiang:2017yfs}, here we just include the corresponding result for comparison.}, including $e^+e^- \to \mu^+ \mu^-, ~Zh, ~W^+W^-, ~ZZ$, and $Z \gamma$. 
 We calculate the deviations of the cross sections of these processes that induced by the NP sector at one-loop level, and give the combined CEPC constraints at 95\% confidence level. The results show that the constraints of these five processes can be complementary to each other in some parameter regions.
 	
 The paper is organized as follows. In Sec. \ref{sec:SDDM}, we give a brief introduction of the SDFDM model, and calculate its one-loop effects on five SM processes $e^+e^- \rightarrow \mu^+ \mu^-, ~Zh, ~W^+W^-, ~ZZ$, and $Z \gamma$ at the CEPC. The combined 95\% confidence results at the CEPC are also shown. In Sec. \ref{sec:DTDM}, we show the results of the DTFDM model. Conclusions and discussions are given in Sec. \ref{sec:con}.
 	
 	\section{Singlet-Doublet fermionic Dark Matter Model}\label{sec:SDDM}
 	
 	\subsection{Model details}
 	In the SDFDM model \cite{Yaguna:2015mva,Calibbi:2015nha,Cai:2016sjz,Xiang:2017yfs,Wang:2018}, the dark sector contains one singlet Weyl spinor $S$ and two doublet Weyl spinors $D_i$ ($i=1,2$) obeying the following $SU(2)_L \times U(1)_Y$ gauge transformations:
 	\begin{equation}\label{model}
 		S\in(\textbf{1}, 0)
 		,~~~~~~
 		D_1=\left( \begin{array}{l}
 			D^0_1 \\
 			D^-_1
 		\end{array} \right)\in(\textbf{2}, -1)
 		,~~~~~~
 		D_2=\left( \begin{array}{l}
 			D^+_2 \\
 			D^0_2
 		\end{array} \right) \in(\textbf{2}, 1).
 	\end{equation}
 	The hypercharge signs of $D_1$ and $D_2$ are opposite, which guarantees that the SDFDM
 model is anomaly free. The gauge invariant Lagrangians are given by
 	\begin{equation}\label{lagrangian_sddm}
 		\begin{split}
 			\mathcal{L}_S & = i S^+ \bar{\sigma^{\mu}}D_{\mu}S - \frac{1}{2}(m_s S^T(-\epsilon)S + h.c.) , \\
 			\mathcal{L}_D & = i D_1^+ \bar{\sigma^{\mu}}D_{\mu}D_1 + i D_2^+ \bar{\sigma^{\mu}}D_{\mu}D_2 + (m_D D^{iT}_1 (-\epsilon)D^j_2 + h.c.) , \\
 			\mathcal{L}_Y & = y_1 S D_1^i H_i - y_2 S D_2^i \tilde{H_i} + h.c.,
 		\end{split}
 	\end{equation}
  where $D_\mu=\partial_\mu -ig t^j A^j_\mu -i\frac{g'}{2}YB_\mu$ is the covariant derivative, $t^j$ are the generators of the corresponding representation of $SU(2)_L$, $Y$ is the hypercharge, $\epsilon \equiv i\sigma^2$, $m_S$ and $m_D$ are the mass parameters of the singlet and doublets, respectively, $y_1$ and $y_2$ are two Yukawa couplings between the dark sector particles and Higgs boson. Consequently, there are  four independent parameters in this model, which are $m_S$, $m_D$, $y_1$, and $y_2$.
 	
 	After EWSB, the Higgs field gets vacuum expectation value $v$ and can be written in the unitary gauge as
 	\begin{equation}
 		H=\frac{1}{\sqrt{2}} \left( \begin{array}{c}
 			0 \\
 			v+h
 		\end{array} \right),~~~~~~~~~~
 		\tilde{H}=\frac{1}{\sqrt{2}} \left( \begin{array}{c}
 			v+h \\
 			0
 		\end{array} \right).
 	\end{equation}
 	Then the Yukawa couplings lead to mixing between the singlet Weyl spinor and doublet Weyl spinors. The mass terms of the dark sector particles are given by
 	\begin{equation}\label{mass_term}
 		\begin{split}
 			\mathcal{L}_m  & = -\frac{1}{2}(S, D_1^0, D_2^0)M_n (-\epsilon) \left( \begin{array}{c}
 				S \\
 				D_1^0 \\
 				D_2^0
 			\end{array} \right) -
 			M_D D_1^- (-\epsilon)D_2^+ + h.c.             \\
 			& = -\frac{1}{2} m_{\chi^0_i} \sum \chi_i^0 (-\epsilon) \chi_i^0 - m_{\chi^{\pm}} \chi^- (-\epsilon) \chi^+ + h.c. ,
 		\end{split}
 	\end{equation}
 	where
 	\begin{equation}\label{mass matrix}
 		M_n = \begin{pmatrix}
 			M_S & \frac{1}{\sqrt{2}} y_1 v & \frac{1}{\sqrt{2}} y_2 v  \\
 			\frac{1}{\sqrt{2}} y_1 v & 0 & -M_D        \\
 			\frac{1}{\sqrt{2}} y_2 v  & -M_D & 0  \\
 		\end{pmatrix}
 	\end{equation}
 is the mass matrix of the neutral particles, $\chi^0_i ~(i = 1,~2,~3)$ and $\chi^\pm$ are the mass eigenstates, $m_{\chi^0_i}$ and $m_{\chi^\pm}$ are the masses of the correspoing mass eignestates. 
 The mass eigenstates of the neutral particles $\chi_i^0$ are connected to the gauge eigenstates through the mixing matrix $\mathcal{N}$. That is to say
  	\begin{equation}		
    \mathcal{N}^T M_n \mathcal{N} = \text{diag}(m_{\chi^0_1},m_{\chi^0_2},m_{\chi^0_3}), ~~~
    	\begin{pmatrix}
 			S \\
 			D_1^0  \\
 			D_2^0   \\
 		\end{pmatrix} = \mathcal{N}
 		\begin{pmatrix}
 			\chi_1^0   \\
 			\chi_2^0   \\
 			\chi_3^0   \\
 		\end{pmatrix}.
 	\end{equation}
For convenience, we adopt the mass orders $m_{\chi^0_1} \leq m_{\chi^0_2} \leq m_{\chi^0_3}$, which can be realized by adjusting the $\mathcal{N}$. Because of the discrete $Z_2$ symmetry, the lightest neutral fermion $\chi^0_1$ is stable and can be regarded as DM candidate. Besides, we can construct 4-component Dirac spinors from 2-component Weyl spinors:
 	\begin{equation}
 		\Psi _i =
 		\begin{pmatrix}
 			\chi_i^0   \\
 			\epsilon (\chi_i^0)^{\dag T} \\
 		\end{pmatrix}, \;\;\;
 		\Psi^+ =
 		\begin{pmatrix}
 			\chi^{+}   \\
 			\epsilon (\chi^-)^{\dag T}
 		\end{pmatrix}.
 	\end{equation}
Then the mass terms as well the interaction terms of SDFDM model can be rewrited as
 	\begin{equation}\label{lagrangian}
 		\begin{split}
 			\mathcal{L}_\text{SDFDM}
 			& = -\frac{1}{2} m_{\chi^0_i} \overline{ \Psi_i} \Psi_i - m_{\chi^\pm} \overline{\Psi^+}\Psi^+ \\
 			& +  -\frac{1}{\sqrt{2}} (y_1 N_{1i}N_{2j} + y_2 N_{1i}N_{3j}) h \overline{ \Psi_i} \Psi_j  + e A_\mu \overline{\Psi^+}\gamma^\mu \Psi^+  + (g\cos\theta - \frac{g\sin^2\theta}{\cos\theta})Z_\mu \overline{\Psi^+}\gamma^\mu \Psi^+           \\
 			& + \frac{g}{4\cos\theta} (N_{2i}^*N_{2j} - N_{3i}^*N_{3j} ) Z_\mu \overline{\Psi_i^0}\gamma^\mu P_L\Psi_j^0  - \frac{g}{4\cos\theta} (N_{2i}^*N_{2j} - N_{3i}^*N_{3j} ) Z_\mu \overline{\Psi_j^0}\gamma^\mu P_R\Psi_i^0     \\
 			&  - \frac{g}{\sqrt{2}}  N_{2i} W^- \overline{\Psi_i^0}\gamma^\mu P_R \Psi^+  + \frac{g}{\sqrt{2}}  N_{3i}^* W^-  \overline{\Psi_i^0}\gamma^\mu P_L \Psi^+  -   \frac{g}{\sqrt{2}}  N_{2i}^* W^+ \overline{\Psi^+}\gamma^\mu P_R \Psi_i^0              \\
 			&  + \frac{g}{\sqrt{2}}  N_{3i} W^+ \overline{\Psi^+}\gamma^\mu P_L \Psi_i^0.
 		\end{split}
 	\end{equation}

\subsection{The detection sensitivity of CEPC}
\label{sec:cepcsen}

 Although the dark sector particle $\chi = \{\chi^0_i,~\chi^\pm\}$ can not be directly produced at the collider when $2m_{\chi}>\sqrt{s}$, it could still affect the cross sections of the SM processes through loop effects, which means the hint of dark sector particles may be revealed through the precision measurements. 
 In this work, we focus on five processes $e^+ e^- \rightarrow \mu^+\mu^-, ~Zh, ~W^+W^-, ~ZZ, $ and $Z\gamma$ at the CEPC, and calculate the deviation of the cross section induced by the dark sector particles from the SM prediction at one-loop level. 
 By utilizing the high precision of the CEPC, these deviations can in turn provide effective constraints on the parameter space of the new physical model. 
 In our calculations, the Packages FeynArts 3.10 \cite{feynarts:mannual}, FormCalc 9.7 \cite{Hahn:1998yk}, and LoopTools 2.15 \cite{Hahn:1998yk} are used to generate Feynman diagrams and derive the numerical results. 
 Note that the whole calculation is performed under the on-shell renormalization scheme.

The cross section at next to leading order with the electroweak correction can be written as
\begin{equation}
    \sigma = \sigma_{\rm{LO}}+\sigma_\text{NLO}^\text{EW}+\sigma_\text{NLO}^\text{Soft},
    \label{eq:oneloop}
\end{equation}
where $\sigma_{\rm{LO}}$ is the tree-level cross section, $\sigma_\text{NLO}^\text{EW}$ is the electroweak correction at one-loop level. $\sigma_\text{NLO}^\text{Soft}$ is the contribution from the soft bremsstrahlung diagrams, which cancels the infrared divergences originate
from the diagrams with the exchange of virtual photons. The deviation of the cross-section from the SM prediction can be expressed as
  \begin{equation}
 \frac{\Delta\sigma}{\sigma_0} = \frac{\mid \sigma_{\rm{SDFDM}}-\sigma_{\rm{SM}} \mid}{\sigma_{\rm{SM}}},
 \end{equation}
 where $\sigma_{\rm{SDFDM}}$ and $\sigma_{\rm{SM}}$ are the one-loop cross sections of the SDFDM model and SM, respectively (see \eqref{eq:oneloop}).
 
 Note that for the process $e^+ e^- \to Z \gamma$, the cross section contains collinear divergence even at the  tree-level. 
 However, considering that the collinear photons can not be detected due to the blind spot of the detector, the collinear divergence can be removed by excluding the collinear photons. 
 Here we calculate the cross section with $|\cos\theta_{\gamma}|<0.99$, where $\theta_{\gamma}$ is the angle between the photon and beam.
 	
The precision of the CEPC is determined by the statistical and systematic uncertainties\footnote{In this analysis, we treat these uncertainties as independent.}.  The statistical uncertainty can be estimated from the predicted event numbers $N$, that is to say, $\sim 1/\sqrt{N}$, while the systematic uncertainties are mainly from the uncertainty of  integrated luminosity and the misidentification of final states. Thereinto, the integrated luminosity uncertainty at the CEPC is $\sim 0.1\%$ \cite{CEPC:CDR}. For the process of $e^+ e^- \to \mu^+ \mu^-$, the misidentification uncertainty can be ignored due to the excellent capability of the muon reconstruction. Note that the leptonic and semileptonic decays of the gauge bosons can also be well reconstructed, while it is not easy to distinguish the two jets from one gauge boson in the pure hadronic decay. For simplicity, in our analysis we only consider the cross sections from leptonic and semileptonic decays, so the reconstruction uncertainty of the gauge bosons can be neglected. Therefore, the systematic uncertainties of $e^+ e^- \to W^+W^-,~ZZ$, and $Z\gamma$ are also only from integrated luminosity. 

According to the number of event in Sec. \ref{sec:intro}, the statistical uncertainties of $e^+ e^- \to \mu^+ \mu^-$,  $W^+W^-$, $ZZ$, and $Z\gamma$ are estimated as 0.032\%, $0.014\%$, $0.059\%$, and  $0.014\%$, respectively.
Combined with the systematic uncertainties, the corresponding precision of these processes are about 
$0.1\%$, $0.1\%$, $0.12\%$, and $0.1\%$. The precision of $e^+ e^- \to Zh$ is given by Ref. \cite{CEPC:CDR}, that is 0.5\%. 
 	
\subsection{Numerical results }
\begin{figure}[htbp]
 		\centering
 		\subfigbottomskip=-100pt
 		\subfigcapskip=-10pt
 		\subfigure[$~e^+e^- \to \mu^+\mu^-$]{\label{fig:sd_eemumu_y1y2} \includegraphics[width=0.45\textwidth]{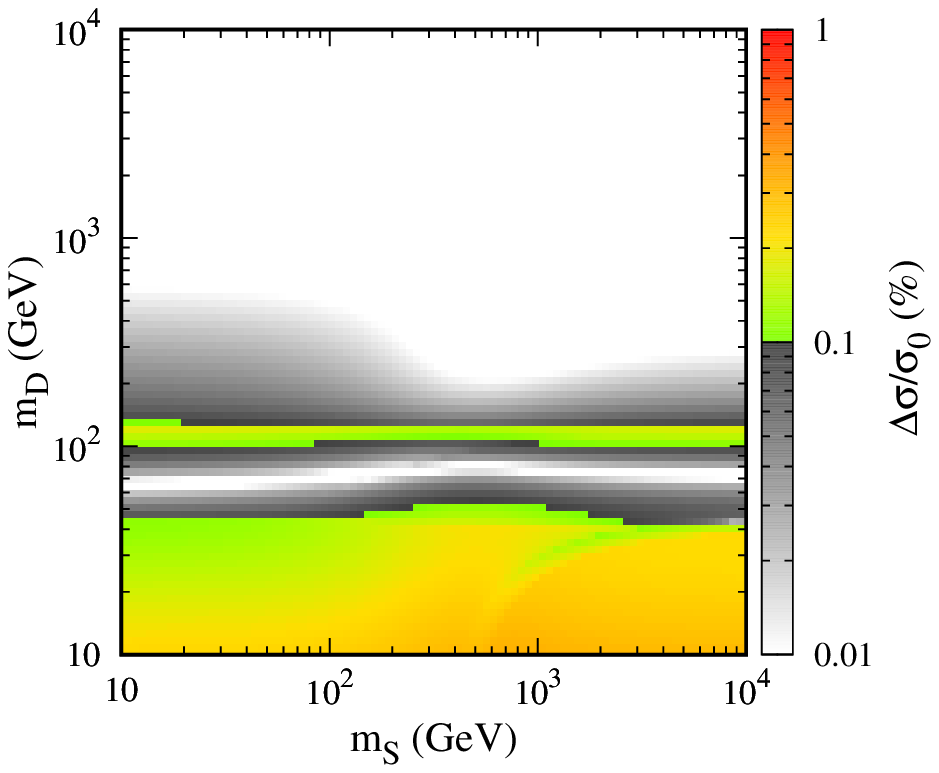}}
 		\subfigure[$~e^+e^- \to Zh$]{\label{fig:sd_eehz_y1y2}     \includegraphics[width=0.45\textwidth]{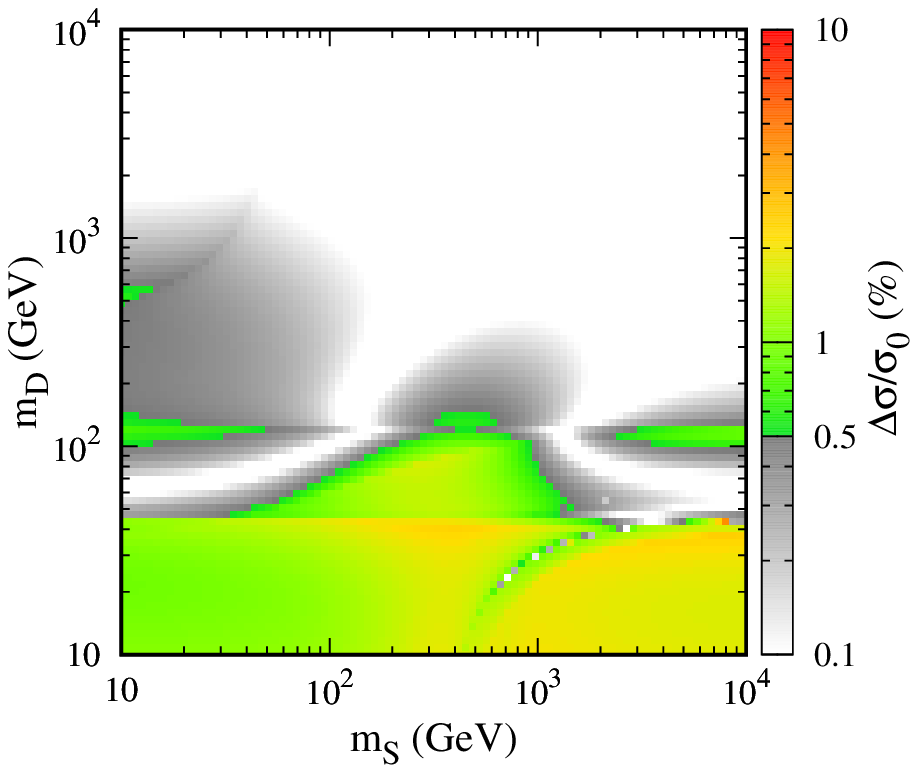}}
 		\subfigure[$~e^+e^- \to W^+W^-$]{\label{fig:sd_eeww_y1y2} \includegraphics[width=0.45\textwidth]{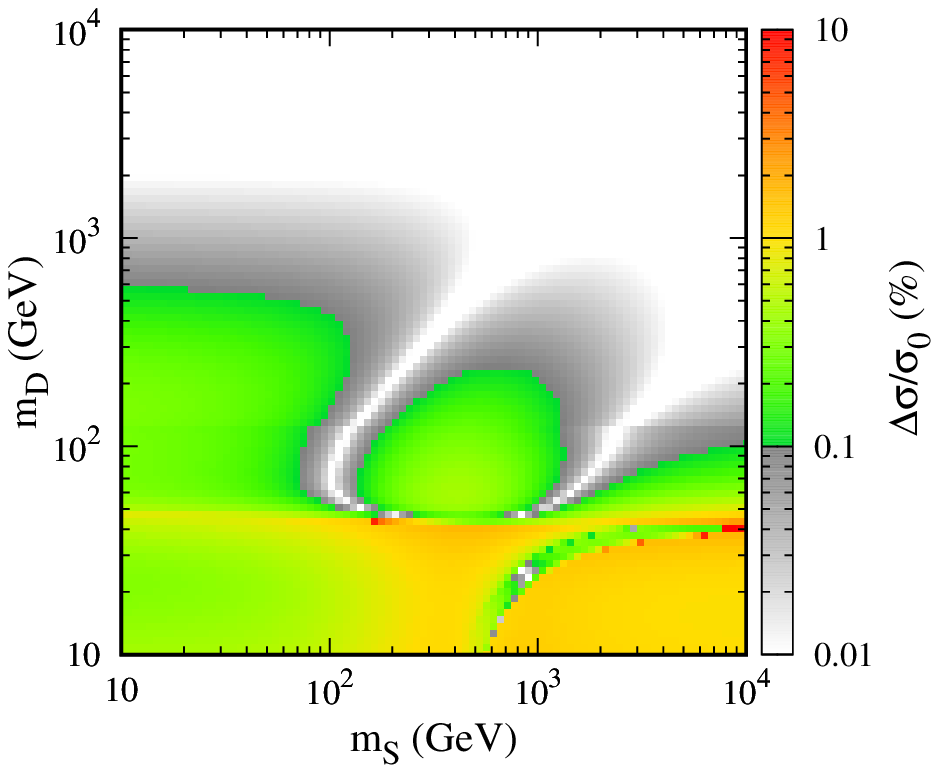}}
 		\subfigure[$~e^+e^- \to ZZ$]{\label{fig:sd_eezz_y1y2} \includegraphics[width=0.45\textwidth]{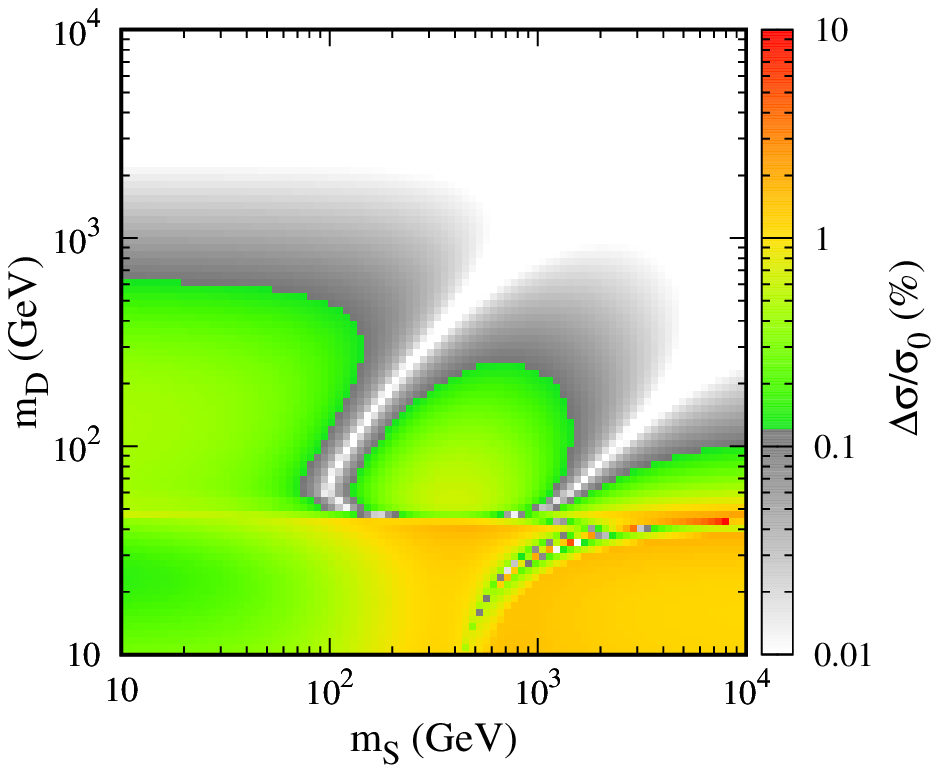}}
 		\subfigure[$~e^+e^- \to Z\gamma$]{\label{fig:sd_eezga_y1y2} \includegraphics[width=0.45\textwidth]{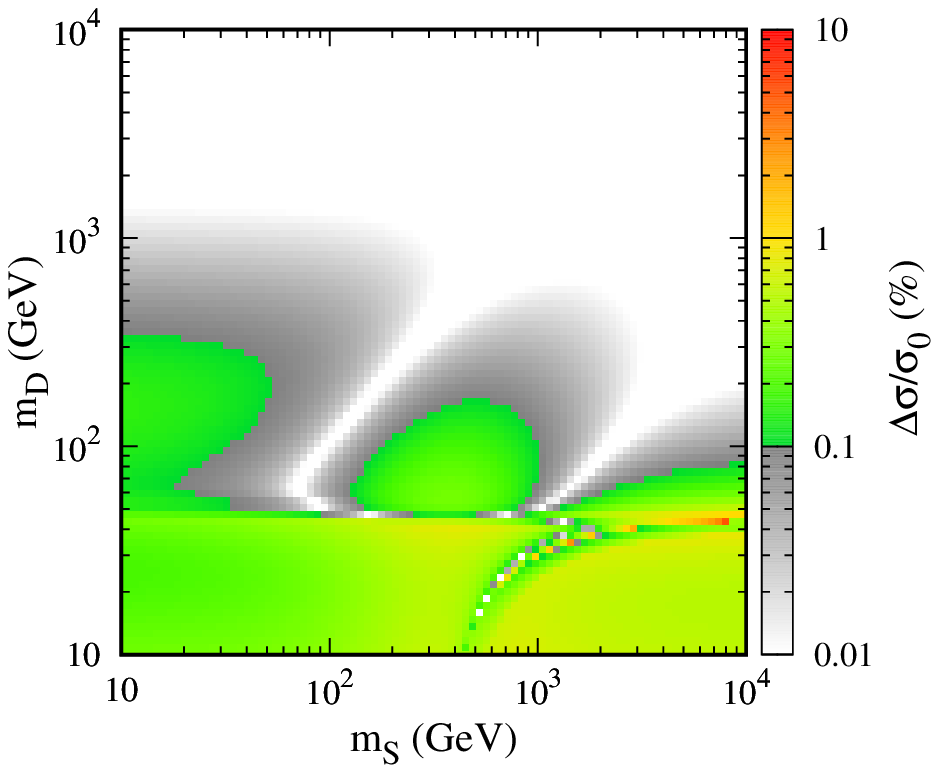}}
 		\subfigure[$~y_1 = 1.0, ~y_2 = 0.5$]{\label{fig:combined_sd_y1y2} \includegraphics[width=0.45\textwidth]{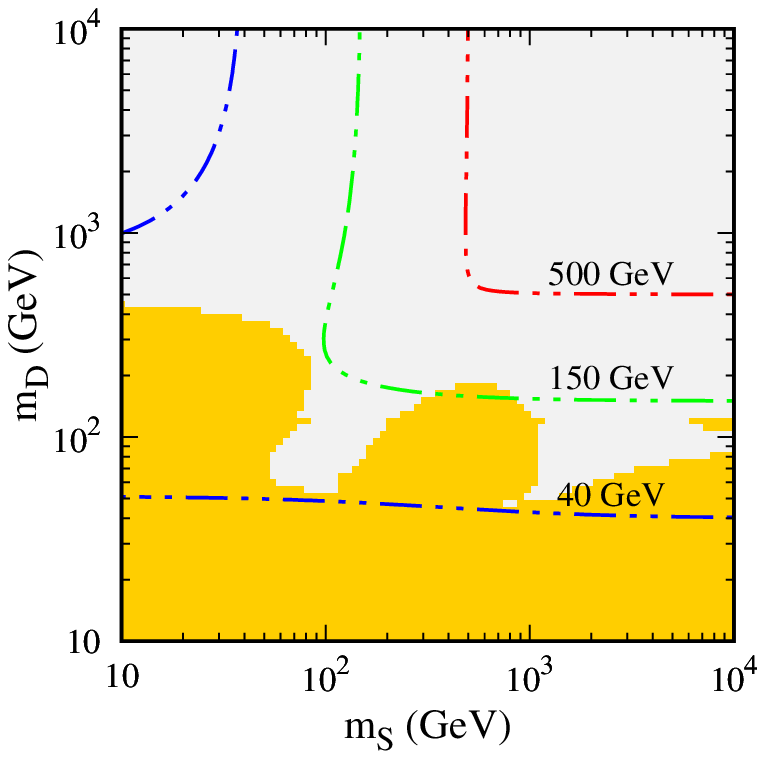}}
 		\caption{Heat maps for the relative deviation of the $e^+e^- \to \mu^+\mu^-,~Zh,~ W^+W^-, ~ZZ$, and $Z\gamma$ are shown in (a) $\sim$ (e). The colored regions indicate that $\Delta \sigma/\sigma_0$ is larger than the expected sensitivities of CEPC.  
 		The combined result of all these channels at 95\% confidence level is shown in (f), where the dot dashed lines represent the mass contour of $\chi^0_1$. Note that all these results are derived with fixed Yukawa couplings of $y_1 = 1.0$ and $y_2 = 0.5$.}
 		\label{fig:sd_y1y2}
 	\end{figure}

\begin{figure}[htbp]
 		\centering
 		\subfigbottomskip=-100pt
 		\subfigcapskip=-10pt
 		\subfigure[$~e^+e^- \to \mu^+\mu^-$]{\label{fig:sd_eemumu_MDMS} \includegraphics[width=0.45\textwidth]{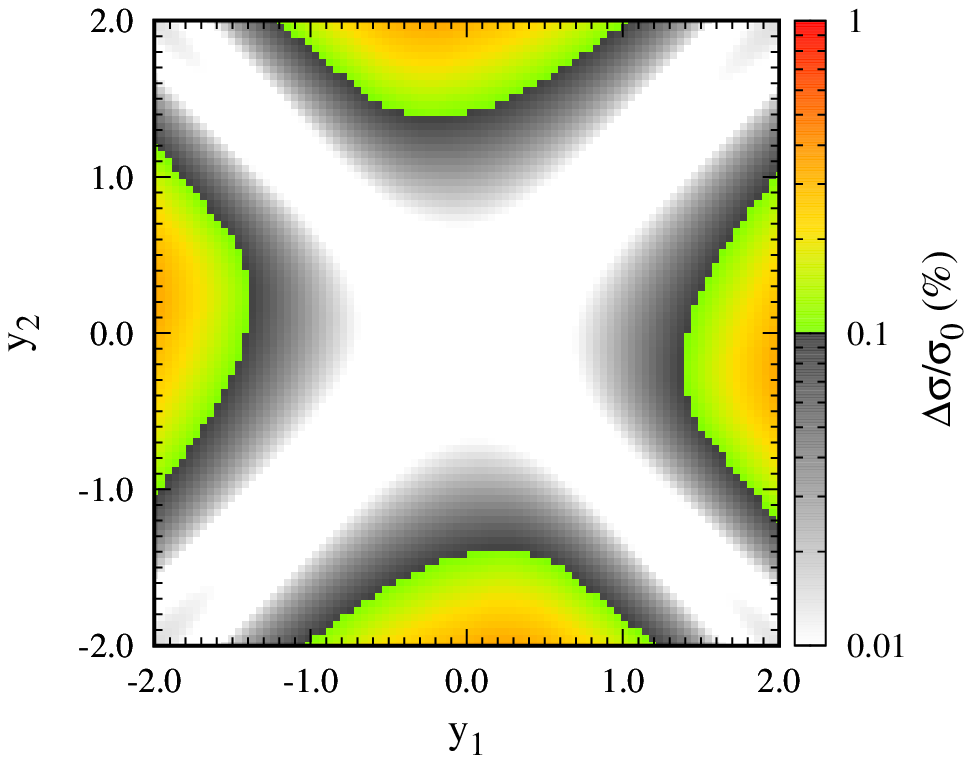}}
 		\subfigure[$~e^+e^- \to Zh$]{\label{fig:sd_eehz_MDMS} \includegraphics[width=0.45\textwidth]{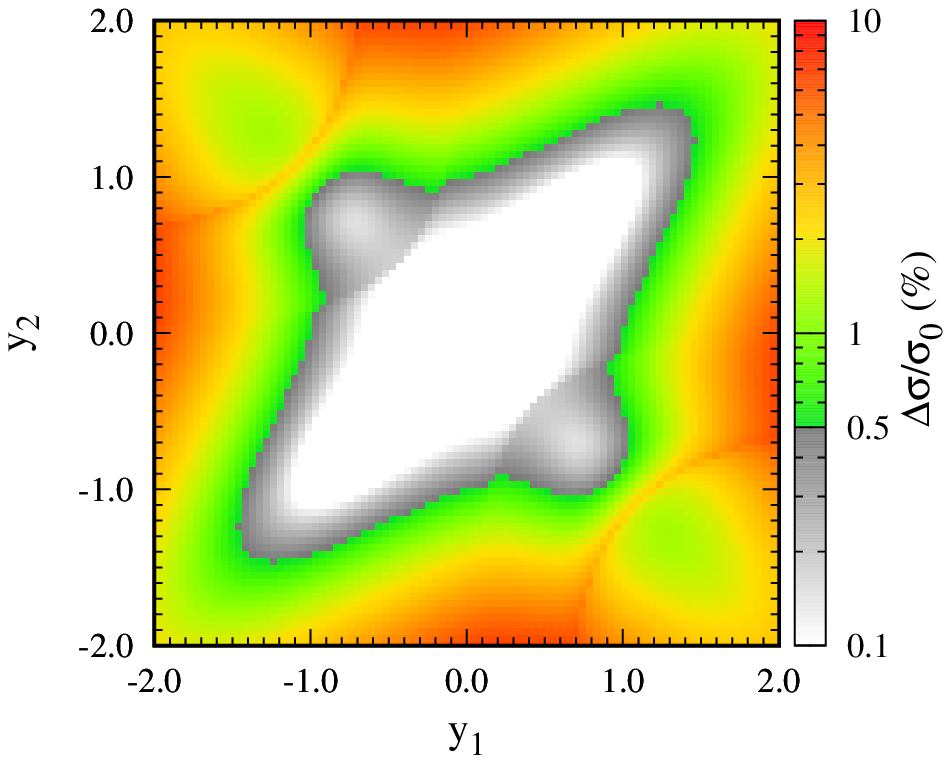}}
 		\subfigure[$~e^+e^- \to W^+W^-$]{\label{fig:sd_eeww_MDMS} \includegraphics[width=0.45\textwidth]{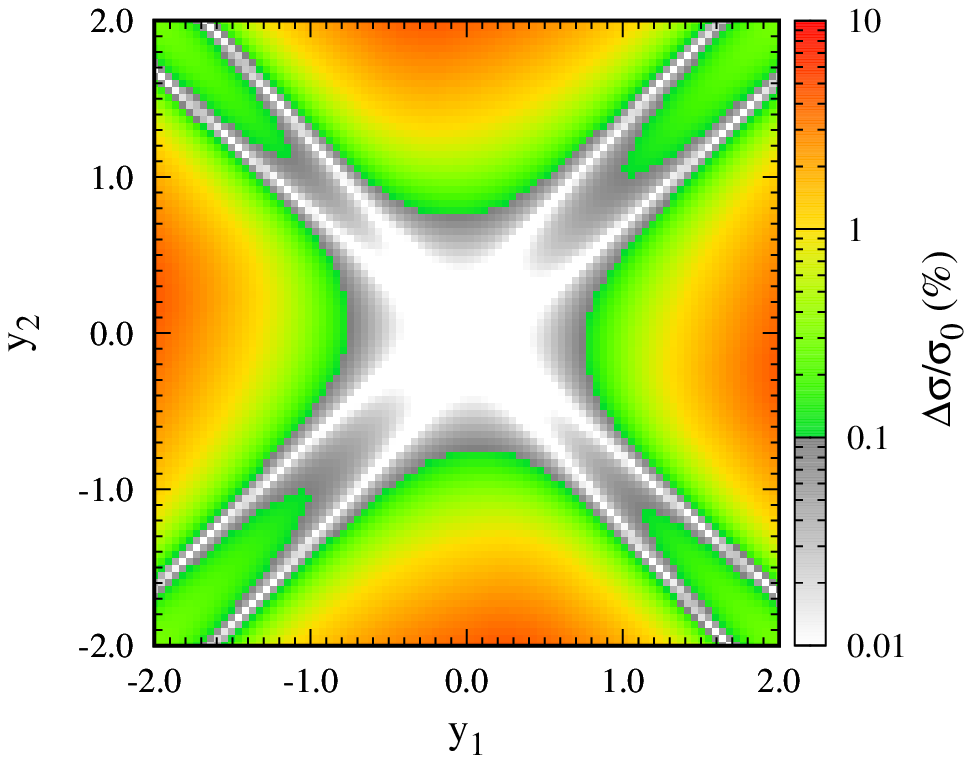}}
 		\subfigure[$~e^+e^- \to ZZ$]{\label{fig:sd_eezz_MDMS} \includegraphics[width=0.45\textwidth]{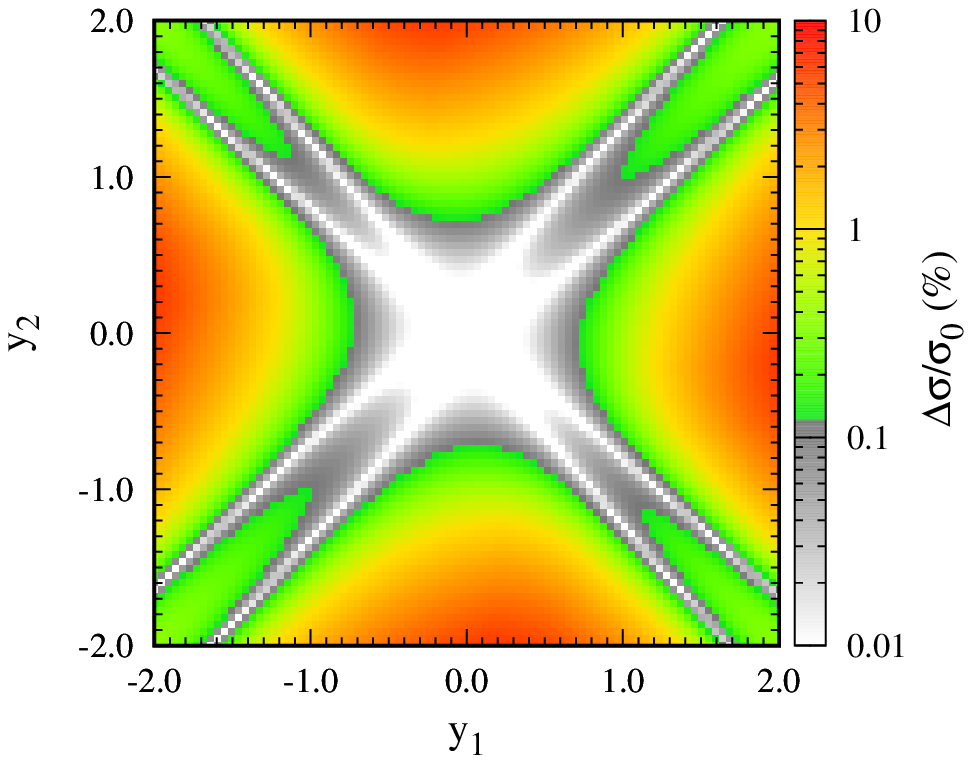}}
 		\subfigure[$~e^+e^- \to Z\gamma$]{\label{fig:sd_eezga_MDMS} ~\includegraphics[width=0.45\textwidth]{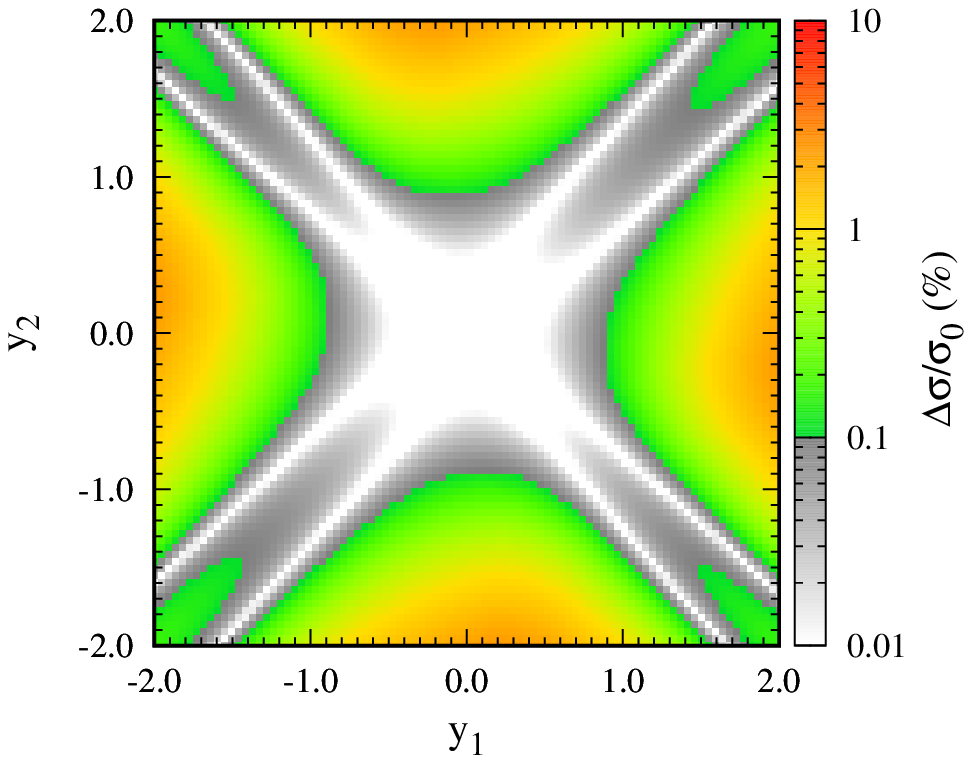}}
 		\subfigure[$~M_S = 100~\text{GeV}, ~M_D = 400~\text{GeV}$]{\label{fig:combined_sd_MSMD} ~~\includegraphics[width=0.45\textwidth]{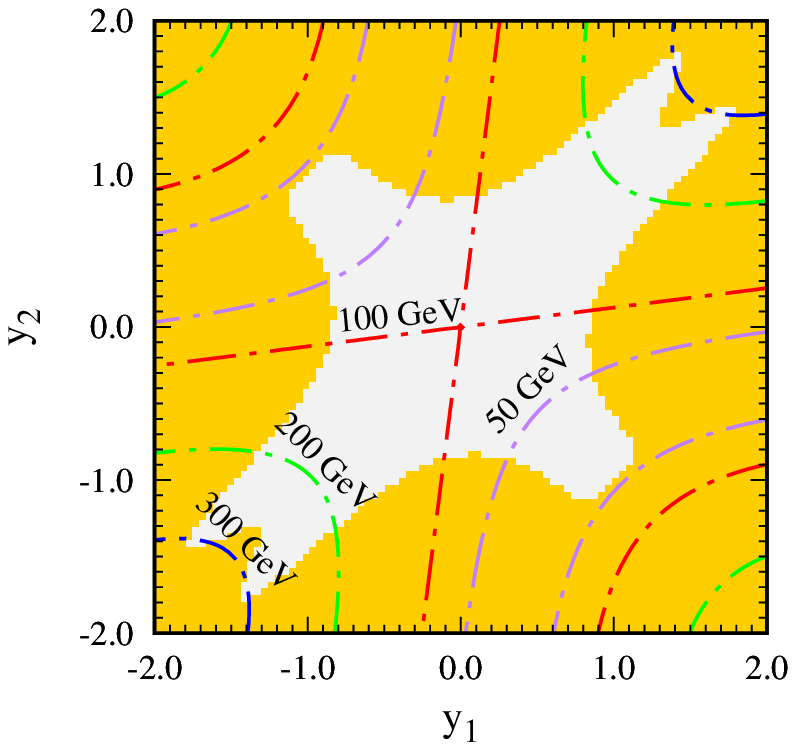}}
 		\caption{Same as Fig. \ref{fig:sd_y1y2}, but in the $y_1-y_2$ plane with the fixed mass parameters of $M_S = 100$ GeV and $M_D = 400$ GeV.}
 		\label{fig:sd_MDMS}
\end{figure}

\begin{figure}[htbp]
 		\centering
 		\subfigbottomskip=-100pt
 		\subfigcapskip=-10pt
 		\subfigure[$~e^+e^- \to \mu^+\mu^-$]{\label{fig:sd_eemumu_MSMD} \includegraphics[width=0.45\textwidth]{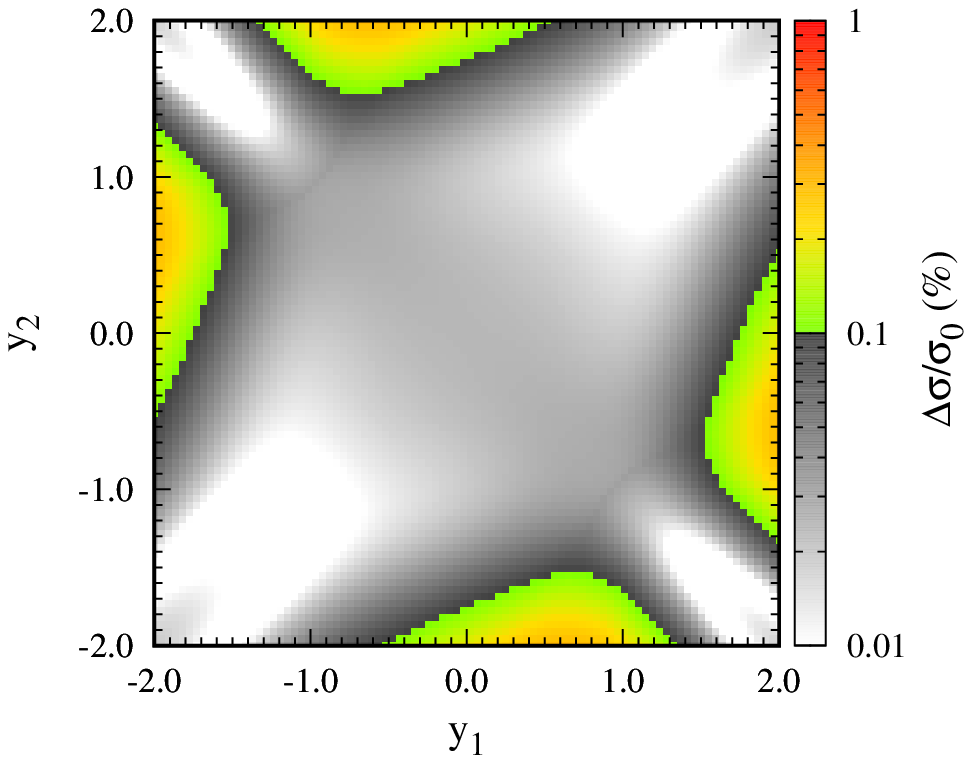}}
 		\subfigure[$~e^+e^- \to Zh$]{\label{fig:sd_eehz_MSMD} \includegraphics[width=0.45\textwidth]{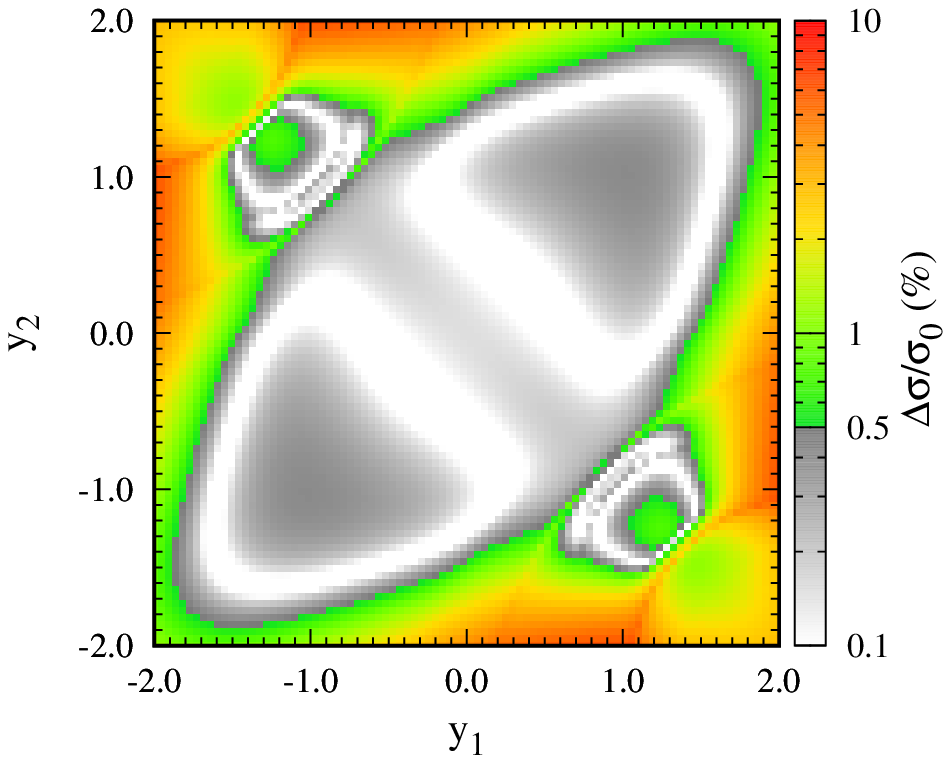}}
 		\subfigure[$~e^+e^- \to W^+W^-$]{\label{fig:sd_eeww_MSMD} \includegraphics[width=0.45\textwidth]{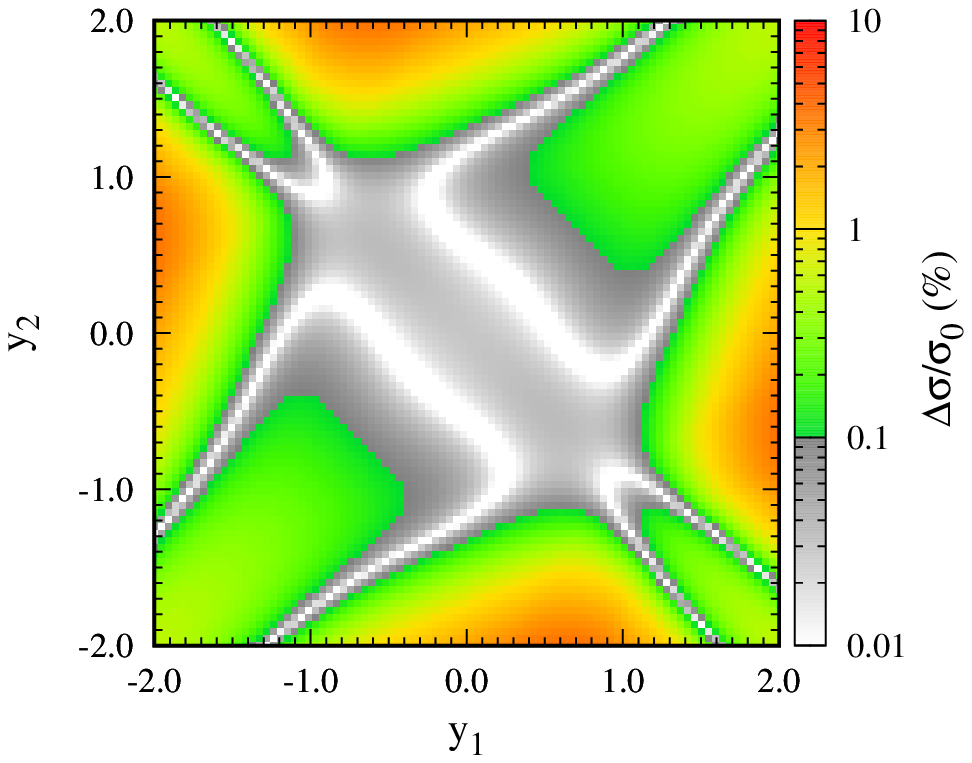}}
 		\subfigure[$~e^+e^- \to ZZ$]{\label{fig:sd_eezz_MSMD} \includegraphics[width=0.45\textwidth]{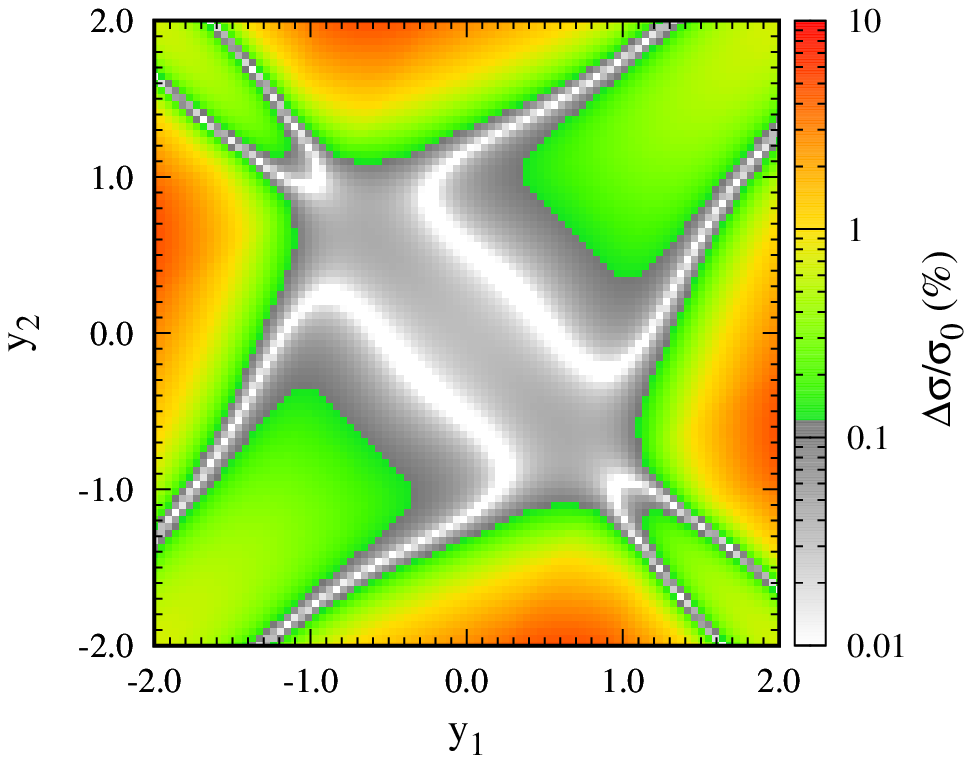}}
 		\subfigure[$~e^+e^- \to Z\gamma$]{\label{fig:sd_eezga_MSMD} ~\includegraphics[width=0.45\textwidth]{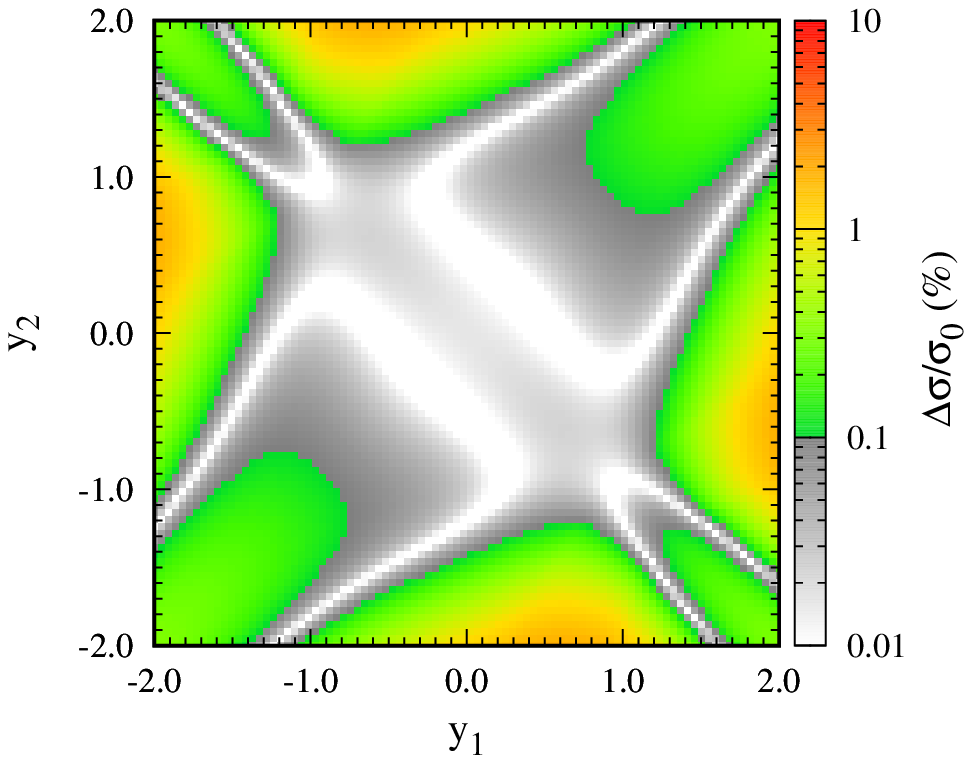}}
 		\subfigure[$~M_S = 400~\text{GeV}, ~M_D = 200~\text{GeV}$]{\label{fig:combined_sd_MDMS} ~\includegraphics[width=0.45\textwidth]{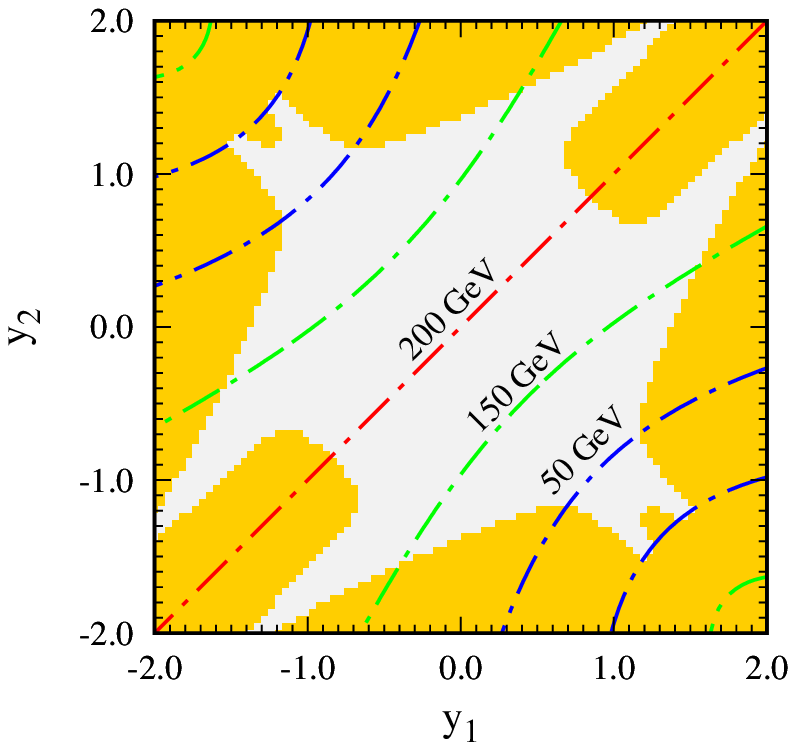}}
 		\caption{Same as Fig. \ref{fig:sd_y1y2}, but in the $y_1-y_2$ plane with the fixed mass parameters of $M_S = 400$ GeV and $M_D = 200$ GeV.}
 		\label{fig:sd_MSMD}
\end{figure}

In this section, we study the capability of the CEPC to probe the parameter space of the SDFDM model. 
Considering that there are four free parameters in the SDFDM model, i.e. $y_1,~y_2,~ m_S,$ and $m_D$, 
we propose to show the calculation results of $\Delta \sigma/\sigma_\text{SM}$ on the Yukawa plane ($y_1-y_2$) or/and mass plane ($m_S-m_D$). 
Here we choose three sets of benchmark parameters: (1) $y_1 = 1.0$ and $y_2 = 0.5$; (2) $M_S = 100$ GeV and $M_D = 400$ GeV; (3) $M_S = 400$ GeV and $M_D = 200$ GeV. 
Considering that the Yukawa interactions induced mass mixing between the singlet and doublets (see Eq. \eqref{lagrangian_sddm}) are  at the order of the EWSB scale $\sim \mathcal{O}$(100) GeV, the benchmark parameter (2) and (3) represent the DM candidate $\chi^0_1$ is roughly singlet-dominated and  doublet-dominated, respectively.
The corresponding results are shown in Fig. \ref{fig:sd_y1y2}, Fig. \ref{fig:sd_MDMS}, and Fig. \ref{fig:sd_MSMD}. 
 
In Fig. \ref{fig:sd_y1y2} we show the results of first set of benchmark parameters, where the Fig. \ref{fig:sd_eemumu_y1y2} $\sim$ Fig. \ref{fig:sd_eezga_y1y2} represent the processes $e^+e^- \to \mu^+ \mu^-$, $Zh$, $W^+W^-$, $ZZ$, and $Z \gamma$, respectively. 
The Fig. \ref{fig:combined_sd_y1y2} represents the combined result of all these channels at 95\% confidence level (see details below). 
The colored parameter regions indicate that $\Delta \sigma/\sigma_0$ is larger than the precision of the CEPC and then could be explored in the future measurement.

In comparison with other channels, the result of $e^+ e^- \to \mu^+ \mu^-$ (see Fig. \ref{fig:sd_eemumu_y1y2}) is relatively simpler. 
We find that the colored regions are strongly dependent on $m_D$ and almost independent of the $m_S$. 
Considering that the dark sector particles in SDFDM model do not interact with lepton, 
the deviation of this process is dominantly induced by the corrections to the propagators, i.e. $Z$ and/or $\gamma$. 
For the charged particle $\chi^\pm$ (pure doublet), its coupling to $Z$ and $\gamma$ are constant (see Eq. \ref{lagrangian}), 
so the corrections induced by $\chi^\pm$ only depend on $m_D$. 
For the neutral particles $\chi^0_i$, 
the results are similar for the case where $m_D \ll m_S$, because in this case the lighter neutral particles $\chi^0_1$ and $\chi^0_2$ are also doublet dominated. 
Our calculation shows that the contribution of SDFDM model to $e^+e^- \to \mu^+ \mu^-$ is dominant by doublets, which is consistent with Fig. \ref{fig:sd_eemumu_y1y2}).
\begin{figure}
 		\centering
 		\includegraphics[width=0.6\textwidth]{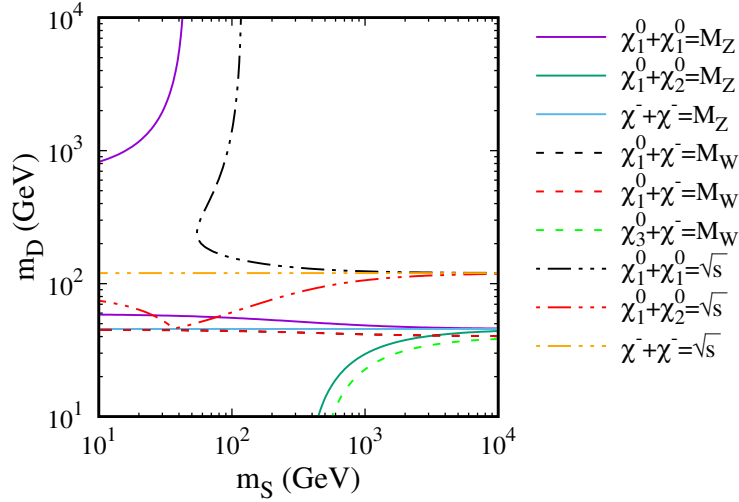}
 		\caption{Contours for the mass threshold conditions with $y_1 = 1.0$ and $y_2 = 0.5$.}.
 		\label{fig:sd_threshold}
\end{figure}
Besides, the corrections from the NP particles in the loop would change dramatically  when some mass conditions are fulfilled. 
These mass threshold effects might partially explain the structures in Fig.~\ref{fig:sd_y1y2}. 

In Fig.~\ref{fig:sd_threshold}, we show some typical mass conditions with $y_1 = 1.0$ and $y_2 = 0.5$, e.g.  $m_{\chi^0_1}+m_{\chi^0_1}(m_{\chi^0_2})=\sqrt{s}$, $2m_{\chi^-}=\sqrt{s}$,  $m_{\chi^0_1}+m_{\chi^0_1}(m_{\chi^0_2})=m_Z$, $2m_{\chi^-}=m_Z$, and $m_{\chi^0_i}+m_{\chi^-}=m_{W^-}$. 
For $e^+e^- \to \mu^+ \mu^-$, we find that the NP correction reaches a maximum value at $m_D \sim 45$ GeV with $2m_{\chi_1^0}\sim m_Z$ and $2m_{\chi_-}\sim m_Z$, and then with the increase of $m_D$, the NP correction quickly decreases to zero and changes to an opposite sign until the next mass threshold appears. This behavior could explain the gray band of Fig. \ref{fig:sd_eemumu_y1y2} within 50 GeV$\lesssim m_D \lesssim 80 ~\rm{GeV}$, where the correction of NP is too small to detect. When the next threshold appears at $m_D \sim 120$ GeV with $2m_{\chi^-}=\sqrt{s}$, the NP correction reaches a minimum value and thus $\Delta \sigma/\sigma_0$ reaches a new maximum value. When $m_D \gtrsim 120$ GeV, the absolute value of the correction is suppressed by the large masses of the NP particles in loops.
Another striking feature is that the results of $e^+ e^- \to W^+W^-$, $ZZ$, and $Z\gamma$ are similar (see Fig. \ref{fig:sd_eeww_y1y2} $\sim$ Fig. \ref{fig:sd_eezga_y1y2}). We find that the contributions from the triangle loops connecting to three gauge bosons for these processes are actually sub-dominant. The main NP corrections are contributed by the diagrams with the insertion of the counter terms into the vertex between the gauge boson and leptons. 

For the second benchmark case, i.e. $m_S = 100~\mathrm{GeV}$ and $m_D = 400~\mathrm{GeV}$, the corresponding results are shown in Fig. \ref{fig:sd_MDMS}.
In this case, the $\chi^0_1$ is dominated by the singlet and is much lighter than the other NP particles. 
Therefore, for the process $e^+ e^- \to \mu^+\mu^-$, the NP correction significantly depends on the $Z\chi^0_1\chi^0_1$ coupling. 
We find that $|y_1| = |y_2|$ leads to a vanishing $Z\chi^0_1\chi^0_1$ coupling and thus the NP correction induced by $\chi^0_1$ is highly suppressed, which rends a non-detectable parameter region around $|y_1| \approx |y_2|$ at CEPC (see Fig. \ref{fig:sd_eemumu_MDMS}).
The structures for the processes of $e^+ e^- \to W^+W^-$, $ZZ$, and $Z\gamma$ are also similar. 
However, the parameter regions with $\mid y_1 \mid \approx \mid y_2 \mid \gtrsim 1$ can be detected at the CEPC, which is different from the process $e^+ e^- \to \mu^+\mu^-$.

For the third benchmark case, i.e. $m_S = 400~\mathrm{GeV}$and $m_D = 200~\mathrm{GeV}$, both the neutral particles $\chi_1$ and $\chi_2$ are doublet-dominated, while $\chi_3$ is singlet-dominated. In this case, $\chi^0_1$ and $\chi^0_2$ would provide a comparable contribution to the deviation. 
However, compare this with the second benchmark case, the mass of $m_{\chi^0_1}$ and $m_{\chi^0_2}$ are heavier, which rends a relatively smaller detectable parameter regions.

From the results shown in subfigures (a)$\sim$(e) of Fig. \ref{fig:sd_y1y2}, Fig. \ref{fig:sd_MDMS}, and Fig.  \ref{fig:sd_MSMD}, we find that for different channels, the parameter space that can be explored by CEPC are actually complementary. 
Therefore, we propose to combine all these five processes and perform a likelihood analysis to get a more efficient constraint on the parameter space \cite{Susufang:2018shg,Harigaya_2015}. We define a $\chi^2$ function as
 	\begin{equation}
 		\chi^2 = \sum_i \frac{(\mu_i^{NP} - \mu_i^{obs})^2}{\sigma^2_{\mu_i}}\simeq \sum_i \frac{(\Delta \sigma/\sigma_0)^2}{\sigma^2_{\mu_i}},
 	\end{equation}
 where $\mu_i^{NP} = \sigma^{NP}_i/\sigma^{SM}_i$ is the ratio of the NLO cross section between NP and SM, $\mu_i^{obs}$ is assumed to be 1, and $\sigma_{\mu_i}$ is the estimated CEPC precision  for the $i$-th process (see Sec. \ref{sec:cepcsen}). For a two-parameter fitting, the corresponding $\delta \chi^2=\chi^2-\chi_{\rm{min}}^2$ at 95\% confidence level is 5.99. 
 The corresponding combined results are shown in Fig. \ref{fig:sd_y1y2}(f), Fig. \ref{fig:sd_MDMS}(f), and Fig. \ref{fig:sd_MSMD}(f), respectively.  For instance, when $y_1$ = 1.0 and $y_2$ = 0.5, the precision measurements at the CEPC could explore the mass of $\chi^0_1$ nearly up to $m_{\chi_1^0} \sim 150~\mathrm{GeV}$ at 95\% confidence level.
 	
\section{Doublet-Triplet Fermionic Dark Matter Model}\label{sec:DTDM}
In the DTFDM model \cite{Dedes:2014hga,Cai:2016sjz,Xiang:2017yfs,Wang:2018}, the dark sector contains one triplet Weyl spinor $T$ and two doublet Weyl spinors $D_i$ ($i=1,2$) obeying the following $SU(2)_L \times U(1)_Y$ gauge transformations: 	
 	\begin{equation}
 		D_1=\left( \begin{array}{l}
 			D^0_1 \\
 			D^-_1
 		\end{array} \right)\in(\textbf{2}, -1)
 		,~~~~~~
 		D_2=\left( \begin{array}{l}
 			D^+_2 \\
 			D^0_2
 		\end{array} \right) \in(\textbf{2}, 1)
 		,~~~~~
 		T = \left( \begin{array}{l}
 			T^+ \\
 			T^0  \\
 			-T^-
 		\end{array} \right) \in(\textbf{3}, 0) .
 	\end{equation}
The gauge invariant Lagrangians of the dark sector particles are given by
 	\begin{equation}
 		\begin{split}
 			\mathcal{L}_T & = i T^+ \bar{\sigma^{\mu}}D_{\mu}T + \frac{1}{2}(m_T T^T(-\epsilon)T + h.c.),  \\
 			\mathcal{L}_D & = i D_1^+ \bar{\sigma^{\mu}}D_{\mu}D_1 + i D_2^+ \bar{\sigma^{\mu}}D_{\mu}D_2 + (m_D D^{iT}_1 (-\epsilon)D^j_2 + h.c.),  \\
 			\mathcal{L}_Y & = y_1 T D_1^i H_i + y_2 T D_2^i \tilde{H_i} + h.c..
 		\end{split}
 	\end{equation}
Similarly, there are also four free parameters, including two mass parameters $m_T$ and $m_D$, and two Yukawa couplings $y_1$ and $y_2$.  
The EWSB leads to the mixing between the triplet and two doublets. In the base of mass eigenstate, we get three neutral Majorana fermions $\chi^0_i~ (i=1,2,3)$ and two charged Dirac fermions $\chi_i^\pm ~(i=1,2)$. The full Lagrangians written in the terms of four-component spinors can be found in Ref. \citep{Xiang:2017yfs}.

\begin{figure}[htbp]
 		\centering
 		\subfigbottomskip=1pt
 		\subfigcapskip=1pt
 		\subfigure[$~y_1 = 1.0, ~y_2 = 0.5~$]{\label{fig:combined_dt_y1y2} \includegraphics[width=0.45\textwidth]{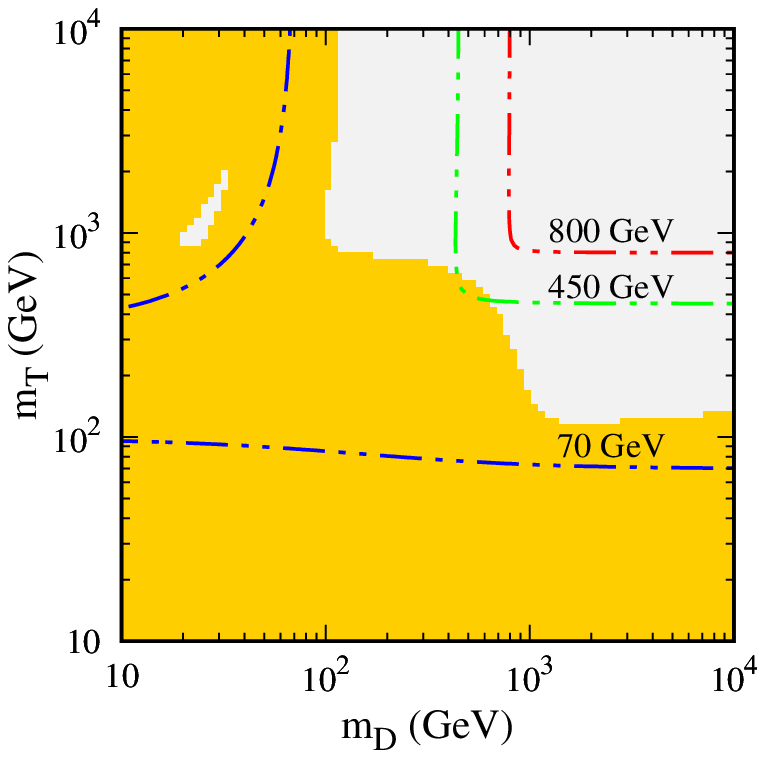}}
 		\subfigure[$~M_D = 100~\text{GeV}, ~M_T = 400~\text{GeV}~$]{\label{fig:combined_dt_MTMD} \includegraphics[width=0.45\textwidth]{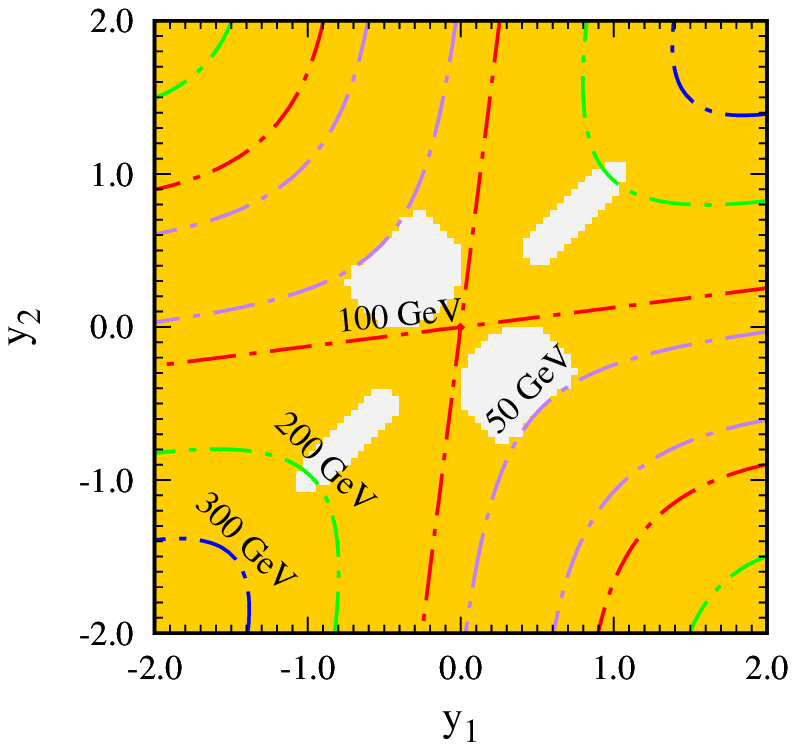}}
 		\subfigure[$~M_D = 400~\text{GeV}, ~M_T = 200~\text{GeV}~$]{\label{fig:combined_dt_MDMT} \includegraphics[width=0.45\textwidth]{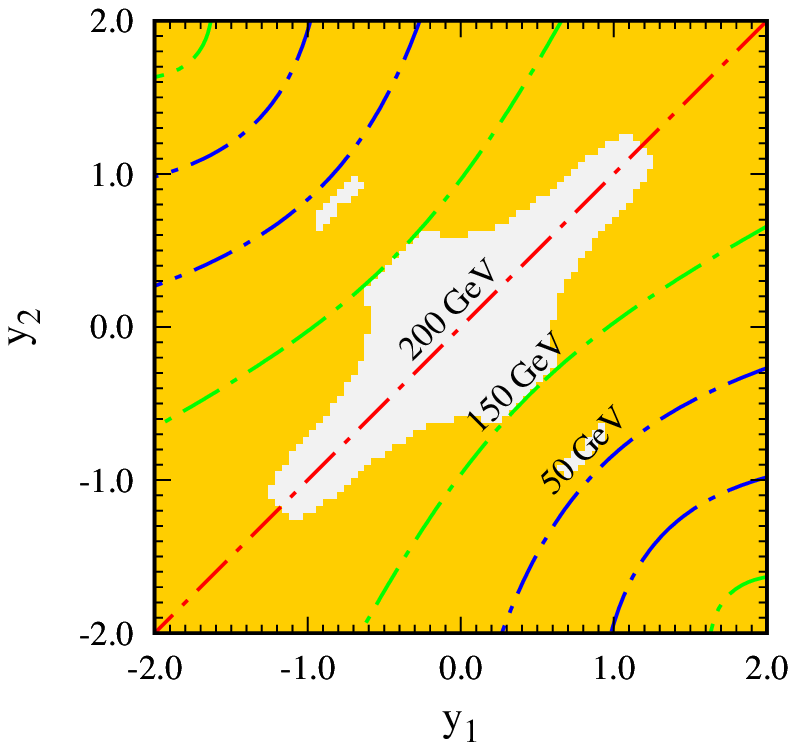}}
 		\caption{Same as subfigure (f) of Fig. \ref{fig:sd_y1y2}, Fig. \ref{fig:sd_MDMS}, and Fig. \ref{fig:sd_MSMD}, but for the DTFDM model.}
 		\label{fig:dt_combined}
\end{figure}
 	
Compared with the SDFDM model, the DTFDM model contains one more charged particles, which would induce relatively larger $\Delta \sigma/\sigma_0$. In Fig.~\ref{fig:dt_combined}, we show the combined constraints from the five processes in the $y_1-y_2$ and $m_D-m_T$ planes, where the yellow regions could be explored by the CEPC at 95\% confidence level.

In Fig.~\ref{fig:combined_dt_y1y2}, we can see that the parameter regions with small $m_D$ or $m_T$ could be explored by CEPC. 
Similar to the results of the SDFDM model for $m_D \lesssim $ 100 GeV, the deviation induced by the DTFDM model can be probed, because two lighter neutral NP particles are doublet dominated and have small masses.
But unlike the results of the SDFDM model for $m_D > 1$ TeV and
$m_s \lesssim $ 100 GeV, the deviation induced by the DTFDM model for $m_D > 1$ TeV and
$m_T \lesssim $ 100 GeV can still be probed. This is because that in this parameter region, except for the lightest neutral NP particles, the lighter charged particles are dominated by the triplet components and have small masses. In Fig. \ref{fig:combined_dt_y1y2} we also show the mass contour of $\chi_1^0$ as Fig. \ref{fig:sd_y1y2}, Fig. \ref{fig:sd_MDMS}, and Fig. \ref{fig:sd_MSMD}. We can see that CEPC have a capability to indirectly detect $m_{\chi^0_1}$ up to 450 GeV.

 
 For the fixed mass parameters $m_D =$100 GeV and $m_T =$400 GeV, almost all the parameter regions with $|y_1|<2$ and $|y_2|<2$ can be excluded by CEPC as shown in Fig.~\ref{fig:combined_dt_MTMD}. This is because that $\chi^0_1$, $\chi^0_2$, and $\chi^\pm_1$ are doublet dominated and provide significant contributions to the NP correction due to small masses. The result for the fixed mass parameters $m_D =$400 GeV and $m_T =$200 GeV is shown in Fig.~\ref{fig:combined_dt_MDMT}. In this case, the triplet dominated $\chi^0_1$ and $\chi^\pm_1$ have relatively large masses in the parameter region with $y_1\approx y_2$, where the couplings $Z\chi^0_i\chi^0_i$ are suppressed. Thus there is a larger region that can not be detected by CEPC compared with the previous case.

\section{ Conclusions }\label{sec:con}

In this work, we study the effects of the EW multiplet fermionic dark matter models through precision measurements at the future electron-positron collider CEPC. As a Higgs factory, CEPC will greatly improve the accuracy of the EW measurements due to its relatively clean environment and large luminosity $\sim5.6$ ab$^{-1}$. 
In particular, the sensitivity of $e^+e^- \to \mu^+\mu^-, ~Zh, ~ZZ, ~W^+W^-$, and $Z\gamma$ can even reach sub-percentage level. 
In this case, any possible deviations from the SM prediction can be regarded as a hint of NP.  

We focus on two fermionic dark matter models, namely SDFDM model and DTFDM model. In these models, two kinds of fermionic multiplets under the $SU(2)_L$ representation are introduced based on SM. 
We calculate the one-loop deviations induced by the NP particles for the above five SM processes and investigate the parameter regions that can be detected by CEPC.
Considering that the results of these channels are complementary for both two models, we adopt a combined analysis and get a more efficient constraint on model parameter space.
For $y_1$ = 1.0 and $y_2$ = 0.5, we find that the CEPC could probe $m_{\chi^0_1}$ up to $ \sim 150~\mathrm{GeV}$ and $\sim 450$ GeV at 95\% confidence level for SDFDM and DTFDM model, respectively.

\section{acknowledgments}
The work of LQG and XJB is supported by the National Natural Science Foundation of China under Grants No. 12175248. The work of JWW is supported by the research grant "the Dark Universe: A Synergic Multi-messenger Approach" number 2017X7X85K under the program PRIN 2017 funded by the Ministero dell'Istruzione, Universit$\grave{a}$ e della Ricerca (MIUR).

\bibliographystyle{utphys}
\bibliography{Ref}

\providecommand{\href}[2]{#2}\begingroup\raggedright\begin{thebibliography}{10}

\bibitem{Aad:2012tfa}
{\bfseries ATLAS} Collaboration, G.~Aad {\em et~al.}, ``{Observation of a new
  particle in the search for the Standard Model Higgs boson with the ATLAS
  detector at the LHC},''
  \href{http://dx.doi.org/10.1016/j.physletb.2012.08.020}{{\em Phys. Lett.}
  {\bfseries B716} (2012) 1--29},
\href{http://arxiv.org/abs/1207.7214}{{\ttfamily arXiv:1207.7214 [hep-ex]}}.

\bibitem{Chatrchyan:2012xdj}
{\bfseries CMS} Collaboration, S.~Chatrchyan {\em et~al.}, ``{Observation of a
  new boson at a mass of 125 GeV with the CMS experiment at the LHC},''
  \href{http://dx.doi.org/10.1016/j.physletb.2012.08.021}{{\em Phys. Lett.}
  {\bfseries B716} (2012) 30--61},
\href{http://arxiv.org/abs/1207.7235}{{\ttfamily arXiv:1207.7235 [hep-ex]}}.

\bibitem{Feng:2010gw}
J.~L. Feng, ``{Dark Matter Candidates from Particle Physics and Methods of
  Detection},''
  \href{http://dx.doi.org/10.1146/annurev-astro-082708-101659}{{\em Ann. Rev.
  Astron. Astrophys.} {\bfseries 48} (2010) 495--545},
\href{http://arxiv.org/abs/1003.0904}{{\ttfamily arXiv:1003.0904
  [astro-ph.CO]}}.

\bibitem{Bertone:2004pz}
G.~Bertone, D.~Hooper, and J.~Silk, ``{Particle dark matter: Evidence,
  candidates and constraints},''
  \href{http://dx.doi.org/10.1016/j.physrep.2004.08.031}{{\em Phys. Rept.}
  {\bfseries 405} (2005) 279--390},
\href{http://arxiv.org/abs/hep-ph/0404175}{{\ttfamily arXiv:hep-ph/0404175
  [hep-ph]}}.

\bibitem{CEPC:CDR}
{\bfseries CEPC Study Group} Collaboration, M.~Dong and G.~Li, ``{CEPC
  Conceptual Design Report: Volume 2 - Physics \& Detector},''
\href{http://arxiv.org/abs/1811.10545}{{\ttfamily arXiv:1811.10545 [hep-ex]}}.

\bibitem{FCC-ee:CDR}
{\bfseries FCC} Collaboration, A.~Abada {\em et~al.}, ``{Future Circular
  Collider},''
CERN-ACC-2018-0057.

\bibitem{ILC:CDR}
H.~Baer, T.~Barklow, K.~Fujii, Y.~Gao, A.~Hoang, S.~Kanemura, J.~List, H.~E.
  Logan, A.~Nomerotski, M.~Perelstein, {\em et~al.}, ``{The International
  Linear Collider Technical Design Report - Volume 2: Physics},''
\href{http://arxiv.org/abs/1306.6352}{{\ttfamily arXiv:1306.6352 [hep-ph]}}.

\bibitem{Cirelli:2005uq}
M.~Cirelli, N.~Fornengo, and A.~Strumia, ``{Minimal dark matter},''
  \href{http://dx.doi.org/10.1016/j.nuclphysb.2006.07.012}{{\em Nucl. Phys.}
  {\bfseries B753} (2006) 178--194},
\href{http://arxiv.org/abs/hep-ph/0512090}{{\ttfamily arXiv:hep-ph/0512090
  [hep-ph]}}.

\bibitem{Yaguna:2015mva}
C.~E. Yaguna, ``{Singlet-Doublet Dirac Dark Matter},''
  \href{http://dx.doi.org/10.1103/PhysRevD.92.115002}{{\em Phys. Rev.}
  {\bfseries D92} (2015) 115002},
\href{http://arxiv.org/abs/1510.06151}{{\ttfamily arXiv:1510.06151 [hep-ph]}}.

\bibitem{Calibbi:2015nha}
L.~Calibbi, A.~Mariotti, and P.~Tziveloglou, ``{Singlet-Doublet Model: Dark
  matter searches and LHC constraints},''
  \href{http://dx.doi.org/10.1007/JHEP10(2015)116}{{\em JHEP} {\bfseries 10}
  (2015) 116},
\href{http://arxiv.org/abs/1505.03867}{{\ttfamily arXiv:1505.03867 [hep-ph]}}.

\bibitem{Cai:2016sjz}
C.~Cai, Z.-H. Yu, and H.-H. Zhang, ``{CEPC Precision of Electroweak Oblique
  Parameters and Weakly Interacting Dark Matter: the Fermionic Case},''
  \href{http://dx.doi.org/10.1016/j.nuclphysb.2017.05.015}{{\em Nucl. Phys.}
  {\bfseries B921} (2017) 181--210},
\href{http://arxiv.org/abs/1611.02186}{{\ttfamily arXiv:1611.02186 [hep-ph]}}.

\bibitem{Xiang:2017yfs}
Q.-F. Xiang, X.-J. Bi, P.-F. Yin, and Z.-H. Yu, ``{Exploring Fermionic Dark
  Matter via Higgs Boson Precision Measurements at the Circular Electron
  Positron Collider},''
  \href{http://dx.doi.org/10.1103/PhysRevD.97.055004}{{\em Phys. Rev.}
  {\bfseries D97} (2018) 055004},
\href{http://arxiv.org/abs/1707.03094}{{\ttfamily arXiv:1707.03094 [hep-ph]}}.

\bibitem{Wang:2018}
J.-W. Wang, X.-J. Bi, P.-F. Yin, and Z.-H. Yu, ``{Impact of Fermionic
  Electroweak Multiplet Dark Matter on Vacuum Stability with One-loop
  Matching},'' \href{http://dx.doi.org/10.1103/PhysRevD.99.055009}{{\em Phys.
  Rev.} {\bfseries D99} (2019) 055009},
\href{http://arxiv.org/abs/1811.08743}{{\ttfamily arXiv:1811.08743 [hep-ph]}}.

\bibitem{Dedes:2014hga}
A.~Dedes and D.~Karamitros, ``{Doublet-Triplet Fermionic Dark Matter},''
  \href{http://dx.doi.org/10.1103/PhysRevD.89.115002}{{\em Phys. Rev.}
  {\bfseries D89} (2014) 115002},
\href{http://arxiv.org/abs/1403.7744}{{\ttfamily arXiv:1403.7744 [hep-ph]}}.

\bibitem{McCullough:2013rea}
M.~McCullough, ``{An Indirect Model-Dependent Probe of the Higgs
  Self-Coupling},'' \href{http://dx.doi.org/10.1103/PhysRevD.90.015001,
  10.1103/PhysRevD.92.039903}{{\em Phys. Rev.} {\bfseries D90} (2014) 015001},
  \href{http://arxiv.org/abs/1312.3322}{{\ttfamily arXiv:1312.3322 [hep-ph]}}.
[Erratum: Phys. Rev.D92,no.3,039903(2015)].

\bibitem{Cao:2014ita}
J.~Cao, Z.~Heng, D.~Li, L.~Shang, and P.~Wu, ``{Higgs-strahlung production
  process $e^{+} e^{-} \to Zh$ at the future Higgs factory in the Minimal
  Dilaton Model},'' \href{http://dx.doi.org/10.1007/JHEP08(2014)138}{{\em JHEP}
  {\bfseries 08} (2014) 138},
\href{http://arxiv.org/abs/1405.4489}{{\ttfamily arXiv:1405.4489 [hep-ph]}}.

\bibitem{Beneke:2014sba}
M.~Beneke, D.~Boito, and Y.-M. Wang, ``{Anomalous Higgs couplings in angular
  asymmetries of $H \to Z\ell^{+} \ell^{-}$ and e$^{+}$ e$^{-} \to HZ$},''
  \href{http://dx.doi.org/10.1007/JHEP11(2014)028}{{\em JHEP} {\bfseries 11}
  (2014) 028},
\href{http://arxiv.org/abs/1406.1361}{{\ttfamily arXiv:1406.1361 [hep-ph]}}.

\bibitem{Shen:2015pha}
C.~Shen and S.-h. Zhu, ``{Anomalous Higgs-top coupling pollution of the triple
  Higgs coupling extraction at a future high-luminosity electron-positron
  collider},'' \href{http://dx.doi.org/10.1103/PhysRevD.92.094001}{{\em Phys.
  Rev.} {\bfseries D92} (2015) 094001},
\href{http://arxiv.org/abs/1504.05626}{{\ttfamily arXiv:1504.05626 [hep-ph]}}.

\bibitem{Huang:2015izx}
F.~P. Huang, P.-H. Gu, P.-F. Yin, Z.-H. Yu, and X.~Zhang, ``{Testing the
  electroweak phase transition and electroweak baryogenesis at the LHC and a
  circular electron-positron collider},''
  \href{http://dx.doi.org/10.1103/PhysRevD.93.103515}{{\em Phys. Rev.}
  {\bfseries D93} (2016) 103515},
\href{http://arxiv.org/abs/1511.03969}{{\ttfamily arXiv:1511.03969 [hep-ph]}}.

\bibitem{Kobakhidze:2016mfx}
A.~Kobakhidze, N.~Liu, L.~Wu, and J.~Yue, ``{Implications of CP-violating
  Top-Higgs Couplings at LHC and Higgs Factories},''
  \href{http://dx.doi.org/10.1103/PhysRevD.95.015016}{{\em Phys. Rev.}
  {\bfseries D95} (2017) 015016},
\href{http://arxiv.org/abs/1610.06676}{{\ttfamily arXiv:1610.06676 [hep-ph]}}.

\bibitem{Wang:2017sxx}
J.-W. Wang, X.-J. Bi, Q.-F. Xiang, P.-F. Yin, and Z.-H. Yu, ``{Exploring
  triplet-quadruplet fermionic dark matter at the LHC and future colliders},''
  \href{http://dx.doi.org/10.1103/PhysRevD.97.035021}{{\em Phys. Rev.}
  {\bfseries D97} (2018) 035021},
\href{http://arxiv.org/abs/1711.05622}{{\ttfamily arXiv:1711.05622 [hep-ph]}}.

\bibitem{Chen:2019pkq}
N.~Chen, T.~Han, S.~Li, S.~Su, W.~Su, and Y.~Wu, ``{Type-I 2HDM under the Higgs
  and Electroweak Precision Measurements},''
  \href{http://arxiv.org/abs/1912.01431}{{\ttfamily arXiv:1912.01431
  [hep-ph]}}.

\bibitem{Susufang:2018shg}
N.~Chen, T.~Han, S.~Su, W.~Su, and Y.~Wu, ``{Type-II 2HDM under the Precision
  Measurements at the $Z$-pole and a Higgs Factory},''
  \href{http://dx.doi.org/10.1007/JHEP03(2019)023}{{\em JHEP} {\bfseries 03}
  (2019) 023},
\href{http://arxiv.org/abs/1808.02037}{{\ttfamily arXiv:1808.02037 [hep-ph]}}.

\bibitem{Cai:2017wdu}
C.~Cai, Z.-H. Yu, and H.-H. Zhang, ``{CEPC Precision of Electroweak Oblique
  Parameters and Weakly Interacting Dark Matter: the Scalar Case},''
  \href{http://dx.doi.org/10.1016/j.nuclphysb.2017.09.007}{{\em Nucl. Phys.}
  {\bfseries B924} (2017) 128--152},
\href{http://arxiv.org/abs/1705.07921}{{\ttfamily arXiv:1705.07921 [hep-ph]}}.

\bibitem{Wulei:2017kgr}
N.~Liu and L.~Wu, ``{An indirect probe of the higgsino world at the CEPC},''
  \href{http://dx.doi.org/10.1140/epjc/s10052-017-5443-z}{{\em Eur. Phys. J.}
  {\bfseries C77} (2017) 868},
\href{http://arxiv.org/abs/1705.02534}{{\ttfamily arXiv:1705.02534 [hep-ph]}}.

\bibitem{Andreev:2012cj}
V.~V. Andreev, G.~Moortgat-Pick, P.~Osland, A.~A. Pankov, and N.~Paver,
  ``{Discriminating Z' from Anomalous Trilinear Gauge Coupling Signatures in
  $e^+ e^- \to W^+ W^-$ at ILC with Polarized Beams},''
  \href{http://dx.doi.org/10.1140/epjc/s10052-012-2147-2}{{\em Eur. Phys. J.}
  {\bfseries C72} (2012) 2147},
\href{http://arxiv.org/abs/1205.0866}{{\ttfamily arXiv:1205.0866 [hep-ph]}}.

\bibitem{Harigaya_2015}
K.~Harigaya, K.~Ichikawa, A.~Kundu, S.~Matsumoto, and S.~Shirai, ``Indirect
  probe of electroweak-interacting particles at future lepton colliders,''
  \href{http://dx.doi.org/10.1007/jhep09(2015)105}{{\em Journal of High Energy
  Physics} {\bfseries 2015} (Sep, 2015) }.

\bibitem{Cao_2016}
Q.-H. Cao, Y.~Li, B.~Yan, Y.~Zhang, and Z.~Zhang, ``Probing dark particles
  indirectly at the CEPC,''
  \href{http://dx.doi.org/10.1016/j.nuclphysb.2016.05.010}{{\em Nuclear Physics
  B} {\bfseries 909} (Aug, 2016) 197–217}.

\bibitem{Gulov:2013kpa}
A.~Gulov, ``{Optimal one-parameter observables for the $Z'$ in $e^+e^- \to
  \mu^+\mu^-$ process},''
  \href{http://dx.doi.org/10.1142/S0217751X14501619}{{\em Int. J. Mod. Phys.}
  {\bfseries A29} (2014) 1450161},
\href{http://arxiv.org/abs/1308.4837}{{\ttfamily arXiv:1308.4837 [hep-ph]}}.

\bibitem{feynarts:mannual}
T.~Hahn, ``{Generating Feynman diagrams and amplitudes with FeynArts 3},''
  \href{http://dx.doi.org/10.1016/S0010-4655(01)00290-9}{{\em Comput. Phys.
  Commun.} {\bfseries 140} (2001) 418--431},
\href{http://arxiv.org/abs/hep-ph/0012260}{{\ttfamily arXiv:hep-ph/0012260
  [hep-ph]}}.

\bibitem{Hahn:1998yk}
T.~Hahn and M.~Perez-Victoria, ``{Automatized one loop calculations in
  four-dimensions and D-dimensions},''
  \href{http://dx.doi.org/10.1016/S0010-4655(98)00173-8}{{\em Comput. Phys.
  Commun.} {\bfseries 118} (1999) 153--165},
\href{http://arxiv.org/abs/hep-ph/9807565}{{\ttfamily arXiv:hep-ph/9807565
  [hep-ph]}}.

\end{thebibliography}\endgroup

\end{document}